\documentclass[prb,aps,superscriptaddress]{revtex4-2}
\usepackage{graphicx}
\usepackage{dcolumn}
\usepackage{bm}
\usepackage{lipsum}
\usepackage{ragged2e}
\usepackage{braket}
\usepackage{bbm}
\usepackage{color}
\usepackage{xcolor}
\usepackage{amsfonts}
\usepackage{amsmath}
\usepackage{amsmath,amssymb,latexsym}
\usepackage{wasysym,stmaryrd,marvosym}
\usepackage{mathrsfs}
\usepackage{mathcomp}
\usepackage{natbib}
\renewcommand{\vec}[1]{\mathbf{#1}}
\usepackage{dsfont}
\providecommand{\keywords}[1]{\textbf{\textit{Index terms---}} #1}
\usepackage[mathlines]{lineno}
\usepackage[colorinlistoftodos]{todonotes}
\usepackage[colorlinks=true, allcolors=blue]{hyperref}

\begin{document}
	
	\newcommand{\cicese}{Centro de Investigación Científica y de Educación Superior de Ensenada, Baja California, Apartado Postal 360, 22860, Ensenada, Baja California, M\'exico.}
	
	\newcommand{\cnyn}{Centro de Nanociencias y Nanotecnolog\'ia,
		Universidad Nacional Aut\'onoma de M\'exico, Apartado Postal 14, 22800, Ensenada, Baja California, M\'exico.}
	
	\title{Linear and nonlinear spin current response of anisotropic spin-orbit coupled systems}
	
	\author{D. Mu\~noz-Santana}
	\affiliation{\cicese}
	\author{Jes\'us A. Maytorena}
	\affiliation{\cnyn}
	
	\date{\today}
	
	\begin{abstract}
		We calculate the linear and the second harmonic (SH)
  spin current response of two anisotropic systems with spin orbit (SO) interaction. 
  The first system is a two-dimensional (2D) electron gas in the presence of Rashba 
  and $k$-linear Dresselhaus SO couplings. The dependence of the 
  anisotropic spin splitting on the sample growth direction introduces an additional 
  path to modify the linear and nonlinear spectra. In particular, vanishing linear and 
  second order spin conductivity tensors are achievable under SU(2) symmetry conditions,
  characterized by a collinear SO vector field. Additional
  conditions under which specific tensor components vanish are posible, without having such collinearity. 
  Thus, a proper choice of the growth direction and SO strengths allows to select the polarization of the linear and SH spin currents according to the direction of flowing. The second system is an anisotropic 2D free electron gas with anisotropic Rashba interaction, which has been employed to study the optical conductivity of 2D puckered structures with anisotropic energy bands. The presence of mass anisotropy and an energy gap open several distinct  scenarios for the allowed optical interband transitions, which manifest in the linear and SH response
  contrastingly. The linear response displays only out-of-plane spin polarized currents, while the SH spin currents flow with spin orientation lying parallel to the plane of the system strictly. The models illustrate the possibility of the 
 nonlinear spin Hall effect in systems with SO interaction, under the presence or 
 absence of time-reversal symmetry. The results suggest different ways 
 to manipulate the linear and nonlinear optical generation of spin currents
 which could find spintronic applications.
  
	\end{abstract}
	
	\keywords{Spin-orbit interaction, linear response, second harmonic generation, anisotropic systems}
	
	\maketitle
	
\section{Introduction}
		
The two-dimensional electron gas (2DEG) in semiconductor heterostructures has maintained its relevance even after almost two decades of constant discovery of new 2D materials. Even before graphene \cite{Novoselov2004}, these quantum wells were the subject of constant research due to their potential applications in the field of spintronics that the presence of spin-orbit (SO) coupling provides \cite{Bandyopadhyay2015,Winkler2003}. The spin Hall effect \cite{Kato2004},
 the current-induced spin polarization \cite{Sih2005}, a nonballistic version of a spin transistor
 \cite{Schliemann2003a,Koo2009},
 and the persistent spin helix state \cite{Koralek2009}
 are remarkable examples of spin-dependent phenomena associated to the 
 SO interaction in 2DEGs. The art of manipulating spin using SO coupling
 gave birth to the field called spin-orbitronics, also dubbed ``Rashba-like 
 physics'' \cite{Manchon2015,Bihlmayer2022,Bercioux2015}, inspired by the Rashba SO coupling, which leads to spin-momentum locking in low dimensional systems lacking inversion symmetry \cite{Winkler2003}.
	
One of the most studied 2DEG systems is the one that includes the Rashba and the  linear Dresselhaus SO contributions. The interplay between these couplings leads to an anisotropic spin-splitting responsible of effects like a characteristic frequency dependence of the charge and spin Hall conductivities \cite{Maytorena2006}, the anisotropy of plasmon dynamics \cite{Badalyan2009}, or persistent spin textures \cite{Bernevig2006,Schliemann2017,Dettwiler2017}, among others. It also allows the possibility of spin-preserving symmetries and associated effects like the infinite spin lifetime due to fixed spin precession axis \cite{Schliemann2017}, the mentioned nonballistic spin-FET \cite{Schliemann2003a}, the suppression of spin beats in oscillations of magnetoresistivity \cite{Averkiev2005}, the vanishing of {\it Zitterbewegung} \cite{Schliemann2006,Biswas2012}, the vanishing of the interband
absorption and spin Hall conductivity \cite{Li2013}, or the cancellation of
plasmon damping \cite{Badalyan2009}, all with potential device applications.
Despite the extensive research, the investigation has usually been 
restricted to quantum wells grown along the [001], [110], or [111] direction \cite{Schliemann2003a,Norman2010,Sherman2006,Glazov2004,Schliemann2017,Wang2014,Belykh2020,Prada2017,Krzyzewska2022,Manchon2009,Bernevig2006,Gambardella2011,Hamamoto2017,Kurebayashi2014,Matos-Abiague2009,Studer2010,Zhou2010}. It is until recently that attention has been paid to a general crystal orientation \cite{Kammermeier2016a,Kozulin2017,Kozulin2019}.
It was found that the recovery of the SU(2) symmetry can be realized for some
growth direction Miller indices, with consequences like the transition from weak anti-localization to weak localization \cite{Kammermeier2016a} or lifetime enhancement
of spin helices \cite{Iizasa2020}.
Recently, the optical generation of dc spin currents by second-order nonlinear 
interactions in 2DEGs with SO coupling has also been explored \cite{Hamamoto2017,Pan2019}.

Another variant of 2D electron systems are new materials presenting
a behavior similar to an electron gas with another type of in-plane anisotropy. These systems, such as phosphorene and group IV metal monochalcogenides, have an orthorhombic lattice and puckered structure with highly anisotropic energy bands, providing novel physical phenomena. Recently, the optical conductivity of such materials was calculated from a model Hamiltonian
consisting of an anisotropic 2DEG \cite{Ahn2021} with an anisotropic Rashba
splitting \cite{Popovic2015}. The optical absorption spectrum revealed a sensitive dependence on the mass anisotropy ratio and on the frequency and direction of the exciting field \cite{Saberi-Pouya2017}.

In this paper, we calculate the linear and the second harmonic (SH) spin current response of two anisotropic systems with SO interaction. 
First, we consider a 2DEG with Rashba and linear-in-momentum Dresselhaus couplings.
We explore how the dependence of the anisotropic spin splitting on the sample growth direction introduces an additional path to modify the linear and nonlinear spectra.
In particular, we find that under SU(2) symmetry conditions,
the linear and second order spin currents vanish. On the other hand, a proper
 choice of the growth direction and SO strengths allows to
select their respective spin polarization according to the direction of flowing.

In reference \cite{Hamamoto2017}, a nonlinear dc spin current generation 
of a Rashba-Dresselhaus coupled 2DEG, grown along the $[001]$ direction,
was investigated as an optical rectification response using a semiclassical
Boltzmann approach in the relaxation-time approximation. Here instead, we focus on
the finite frequency SH response in the clean limit of the system, and
for arbitrary crystallographic direction of growth.

The second system is the same mentioned above
\cite{Saberi-Pouya2017} to study the optical response of 2D 
SO coupled puckered structures with mass anisotropy. We extend it by
introducing an small energy gap in the Hamiltonian to represent a
possible splitting of the conduction band \cite{Saberi-Pouya2018}. This allows us to switch between a model which preserves the time-reversal symmetry (TRS) and one which breaks it. We find that the dependence of the spectrum of 
allowed interband transitions on the mass ratio and energy gap, leads to
out-of-plane spin polarized linear currents, while the SH spin currents
flow with the spin oriented parallel to the plane of the system strictly.
Our results suggest different ways to manipulate the polarization and direction of the optically induced linear and SH spin current responses.

The paper is organized as follows. In section II the Kubo expressions for the
dynamical linear and second-order spin current conductivity tensors of a generic
two-band model are presented. In sections III we evaluate the formulae obtained
in section II for the 2DEG with generic linear-in-momentum
SO interaction and discuss the spectral features of the optical spin current
response at the fundamental and SH frequencies. To this end, we first calculate the joint density of states (JDOS) and explain the origin of its van Hove singularities. 
Similarly, in section IV we discuss the spectral characteristics of the JDOS,
and calculate the linear and SH spin current conductivities for the gapped anisotropic Rashba model. The conclusions are summarized in section V.	
	
	\section{First and second order Kubo formulae} 
 
	In this section we obtain the conductivity tensors characterizing the spin current response which depend linearly and 
	quadratically on an externally applied electric field.
	We shall consider a generic two band model in 2D described by the  momentum-space Hamiltonian $H({\bf k})=\varepsilon_0({\bf k})\mathbb{I}+\boldsymbol{\sigma}\cdot {\bf d}({\bf k})$, with energy spectrum	$\varepsilon_{\lambda}({\bf k})=\varepsilon_0({\bf k}) +\lambda d({\bf k})$, where ${\bf k}$ is the in-plane electron wave vector ${\bf k}=k_x{\bf\hat{x}}+k_y{\bf\hat{y}}=k(\cos\theta{\bf\hat{x}}+\sin\theta{\bf\hat{y}})$,
	${\bf d}({\bf k})=d_x({\bf k}){\bf\hat{x}}+d_y({\bf k}){\bf\hat{y}}+d_z({\bf k}){\bf\hat{z}}$, $d({\bf k})=|{\bf d}({\bf k})|$,
	 $\boldsymbol{\sigma}$ is the vector of Pauli matrices defined in the spin space, $\mathbb{I}$ is the $2\times 2$ unit matrix, and
	$\lambda=\pm$ specifies the helicity of the states in the upper ($+$) and lower ($-$) part of the spectrum.	The eigenstates are 
	$|\lambda,{\bf k}\rangle=[N_{\lambda},\, \lambda N_{-\lambda}\exp{(i\phi)}]^T$, with $[N_{\lambda}({\bf k})]^2=(d+\lambda d_z)/2d$ and $\tan\phi({\bf k})=d_y/d_x$.
	In what follows we simplify the notation and write $|\lambda\rangle$ for this states. 
	
	Within the Kubo formalism, the linearly induced spin current is given by
	\begin{align}\label{KUBO1}
		\braket{{\cal J}_i^{\ell}(t)}^{(1)} =  \frac{1}{i\hbar} \int^{t}_{-\infty} \! dt' \sum_{\lambda,{\bf k}} f(\varepsilon_{\lambda}({\bf k})) \langle  \lambda | [ \hat{{\cal J}}^{\ell}_i(t), \hat{H}'(t')] | \lambda \rangle,
	\end{align}
	where all operators are in the interaction picture with respect to the unperturbed Hamiltonian.
	Here,  $\mathcal{\hat{J}}^{\ell}_{i} = \frac{\hbar}{4}(\sigma_{\ell}\hat{v}_{i} + \hat{v}_{i}\sigma_{\ell})$ is the operator for the $\ell$-polarized spin current flowing in the $i$-direction,
	with $\hbar\hat{v}_i=\partial H/\partial k_i$ defining the velocity operator component $\hat{v}_i$, ($i=x,y; \,\ell=x,y,z$). The operator $\hat{H}'(t)=(e/c)\hat{v}_iA_i(t)$ (sum over repeated index is assumed hereafter) contains
	the interaction of the electrons with a spatially
	homogeneous vector potential ${\bf A}(t)$, ($e>0$).
	The factor $f_{\lambda}=f(\varepsilon_{\lambda}({\bf k}))$ is the Fermi-Dirac occupation number of the band $\varepsilon_{\lambda}({\bf k})$.
	
	Assuming an applied uniform electric field of the form $ E_i(t) = E_i(\omega)e^{-i\omega t} + c.c. =-(1/c)\partial A_i/\partial t$, the time integral leads to a first order induced spin current $ \langle \mathcal{J}^{\ell}_{i}(t) \rangle = e^{-i\omega t} \sigma^{\ell}_{ij}(\omega) E_{j}(\omega) +c.c.$,  with the spin conductivity tensor given by
	\begin{equation}\label{SPINCOND1}
		\sigma^{\ell}_{ij} (\omega) = \frac{ie}{\hbar\tilde{\omega}} \sum_{\lambda\lambda'} \sum_{\bf k} \left( f_{\lambda'} - f_{\lambda} \right) \frac{\langle \lambda | \mathcal{\hat{J}}^{\ell}_{i} | \lambda' \rangle \langle \lambda' | \hat{v}_j | \lambda \rangle }{\tilde{\omega} - \omega_{\lambda'\lambda}} ,
	\end{equation} 	
	where $\tilde{\omega}=\omega+i0^+$, $i,j=x,y$ and $\ell=x,y,z$. Explicitly, we have

	\begin{equation}\label{SPINCONDLINEAR}
		\sigma^{\ell}_{ij}(\omega) = \frac{ie}{2\hbar\omega} \int\frac{d^2k}{(2\pi)^2}
		\left( f_{-} - f_{+} \right) \frac{\partial \varepsilon_{0}}{\partial k_i} \frac{\partial d_p}{\partial k_j} \left\{ \frac{M^{+}_{\ell p}}{\hbar\tilde{\omega} + 2d} - \frac{M^{+}_{p\ell}}{\hbar\tilde{\omega} - 2d} \right\},
	\end{equation}
	where $ M^{\lambda}_{ij}({\bf k}) = \bra{\lambda} \sigma_{i} \ket{-\lambda} \bra{-\lambda} \sigma_j \ket{\lambda} =
	(d^2\delta_{ij}-d_id_j)/d^2+i\lambda\epsilon_{ijk}d_k/d$. 
	
	At second order, the Kubo formula for the induced spin current reads as
	\begin{align}\label{KUBO2}
		\braket{{\cal J}_i^{\ell}(t)}^{(2)} =  \frac{1}{(i\hbar)^2} \int^{t}_{-\infty} \! dt' \int^{t'}_{-\infty} \! dt'' \sum_{\lambda,{\bf k}} f(\varepsilon_{\lambda}({\bf k})) \langle  \lambda| [[ \hat{{\cal J}}^{\ell}_i(t), \hat{H}'(t')], \hat{H}'(t'')]| \lambda \rangle.
	\end{align}
	For this case we assume an applied electric field $ E_i(t) = E_i(\omega_1)e^{-i\omega_1 t} + E_i(\omega_2)e^{-i\omega_2 t} + c.c. $ and focus on the sum frequency (SF) contribution proportional to $E_j(t')E_l(t'')$. The time integrals
	leads to a second order SF spin current 
	$ \langle \mathcal{J}^{\ell}_{i}(t) \rangle_{SF} = e^{-i(\omega_1 + \omega_2)t} \sigma^{\ell,SF}_{ijl}(\omega_1,\omega_2) E_{j}(\omega_1) E_{l}(\omega_2) +c.c.$, with the SF spin conductivity
	tensor given by
	\begin{equation}\label{SPINCOND}
		\sigma^{\ell,SF}_{ijl} (\omega_1,\omega_2) = \frac{e^2}{\hbar^{2}\tilde{\omega}_1\tilde{\omega}_2} \sum_{\wp} \sum_{\lambda\lambda'\lambda''} \int\frac{d^2k}{(2\pi)^2}
		\frac{\big\langle \lambda \big| \mathcal{\hat{J}}^{\ell}_{i} \big| \lambda' \big\rangle \big\langle \lambda' \big| \hat{v}_j \big| \lambda'' \big\rangle \big\langle \lambda'' \big| \hat{v}_l \big| \lambda \big\rangle}{\tilde{\omega}_3 - \omega_{\lambda'\lambda}} \left( \frac{f_{\lambda''} - f_{\lambda}}{\tilde{\omega}_2 - \omega_{\lambda''\lambda}} - \frac{f_{\lambda'} - f_{\lambda''}}{\tilde{\omega}_1 - \omega_{\lambda'\lambda''}} \right),
	\end{equation}
	where $i,j,l=x,y$ and $\ell=x,y,z$. Here $\tilde{\omega}_i=\omega_i+i0^+$, $i=1,2,3$, $\omega_3=\omega_1+\omega_2$, and
	$ \sum_{\wp} $ stands for the intrinsic permutation symmetry $(j,\omega_1)\leftrightarrow (l,\omega_2)$.
	We shall be focused on the second harmonic response,
	$\sigma^{\ell,(2\omega)}_{ijl}(\omega) = \sigma^{\ell,SF}_{ijl}(\omega,\omega)$, 
	\begin{equation}\label{2OMEGATWOLEVEL}
		\sigma^{\ell,(2\omega)}_{ijl}(\omega) = \frac{e^2}{(\hbar\tilde{\omega})^2} \sum_{\wp} \int\frac{d^2k}{(2\pi)^2} (f_{-} - f_{+}) \frac{1}{d} \frac{\partial\varepsilon_{0}}{\partial k_i} \frac{\partial d_p}{\partial k_j} \frac{\partial d_q}{\partial k_l}\! \left[ \! \frac{d_{\ell}M^+_{pq}}{(\hbar\tilde{\omega})^2 - (2d)^2} -  \frac{d_pM^+_{\ell q}}{(\hbar\tilde{\omega} + 2d)(2\hbar\tilde{\omega} + 2d)} - \frac{d_pM^+_{q\ell}}{(\hbar\tilde{\omega} - 2d)(2\hbar\tilde{\omega} - 2d)}  \! \right].
	\end{equation}
	
	In the following, we use expressions (\ref{SPINCONDLINEAR}) and (\ref{2OMEGATWOLEVEL}) to study the spin current response of two anisotropic SO coupled systems.
	
	\section{2DEG with Rashba and Dresselhaus[$hkl$] SO coupling}
	
	\subsection{Joint density of states}
	
	We consider a 2DEG with an arbitrary crystal orientation defined by the unit normal
	${\bf\hat{n}}=n_x{\bf\hat{x}}+n_y{\bf\hat{y}}+
	n_x{\bf\hat{z}}$, with underlying basis vectors
	${\bf\hat{x}}, {\bf\hat{y}}, {\bf\hat{z}},$ pointing along the crystal axes $[100], [010], [001]$, respectively. The system is in the presence of
	Rashba and linear-in-$k$ Dresselhaus SO couplings.
	The corresponding Hamiltonian is determined by the
	function $\varepsilon_0(k_x,k_y)=\hbar^2 k^2/2m$ and the SO vector field $d_i(k_x,k_y)=\mu_{i\nu}k_{\nu}\,(i=x,y,z; \nu=x,y)$. 
	For a given normal ${\bf\hat{n}}$, the $xyz$ coordinate system can be rotated in order to obtain
	a new $x'y'z'$ coordinate system, with the $z'$-axis pointing along the direction of ${\bf\hat{n}}$,
	and orthonormal basis $\hat{{\bf l}},\hat{{\bf m}},\hat{{\bf n}}$, $\hat{{\bf l}}={\bf \hat{m}}\times{\bf \hat{n}}$. 
	Thus, for each orientation the matrix of SO material parameters $\mu_{ij}$
	must be understood as referred to
	the new coordinates, as well as the condition
	${\bf k}\cdot{\bf\hat{n}}=0$, which is transformed to the condition $k'_z=0$. We shall use the symbol
	R+D$[hkl]$ to indicate the Rashba (R) and Dresselhaus (D) SO couplings when the	sample is grown along the crystallographic direction $[hkl]$.
	
\begin{figure}
\centering
\includegraphics[scale=0.35]{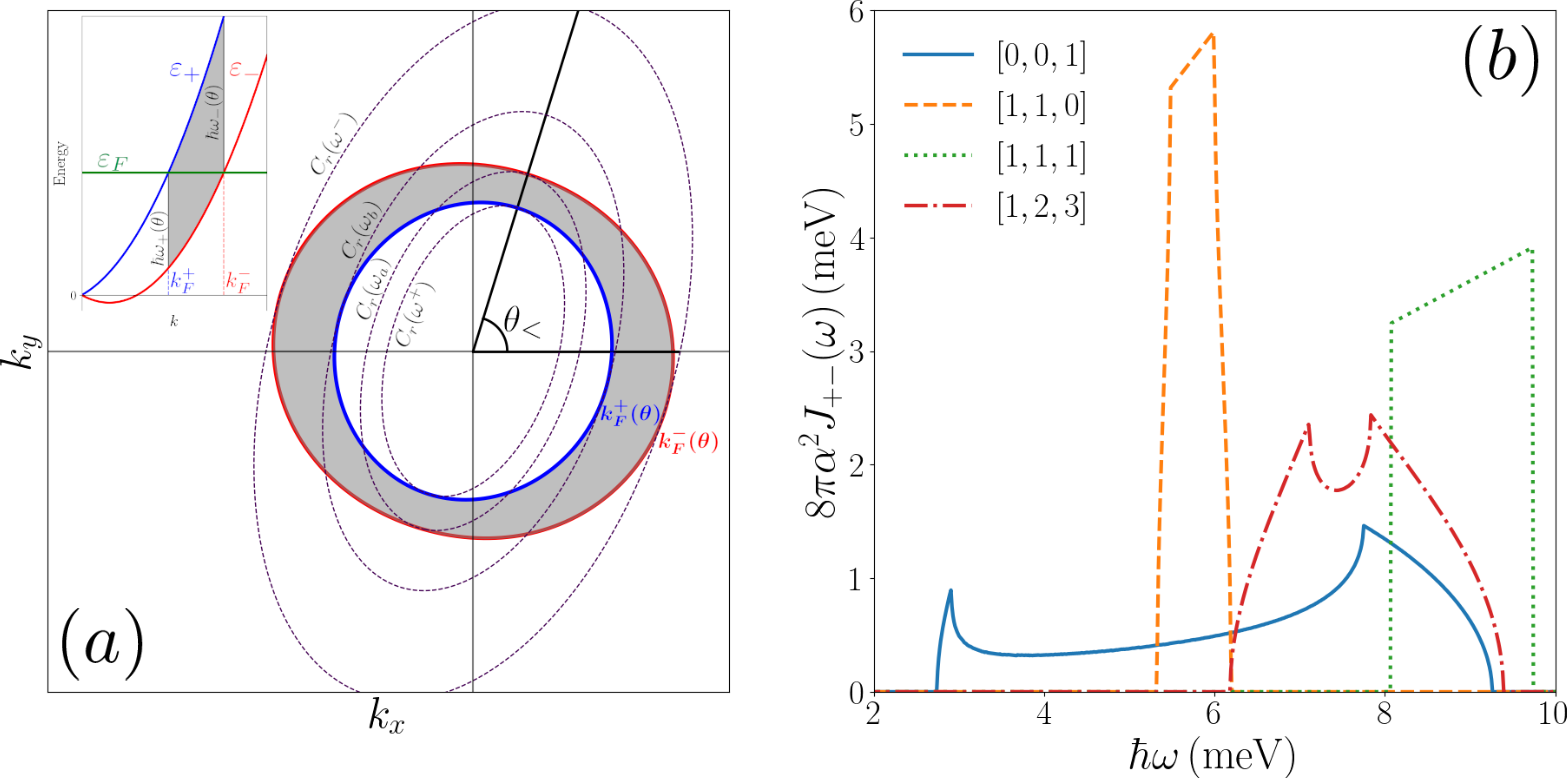}
\caption{(a) Fermi Contours $k_F^{\pm}$ and resonance curves $ C_r(\omega) $ for the critical frequencies $\omega^{\pm}, \omega_a,\omega_b$ ($ \omega^{+}<\omega_{a} < \omega_{b} < \omega^{-} $) of a R+D[$123$] system. At temperature $T=0$, the only states involved in vertical transitions between the bands $ \varepsilon_{\lambda}(\vec{k}) $ (gray area) are those satisfying $ k^{+}_{F}(\theta) < k < k^{-}_{F}(\theta) $, for which $ \varepsilon_{-}(\vec{k}) \leq \varepsilon_F \leq \varepsilon_{+}(\vec{k}) $ (see inset). (b) The JDOS spectrum for several crystal orientations; the parameters used are $ \alpha= 160\,\operatorname{meV} $ \r{A}, $\gamma k^{2}_{n} = 0.5\alpha$, and $n= 5 \times 10^{11}\,  \mathrm{cm}^{-2}$, $m=0.05m_{0}$ for the electron density and effective mass.}
\label{fig:LEVELSANDCONTOURS}
\end{figure}

	The $k$-space available for vertical transitions is determined by the condition $\varepsilon_{-}(\vec{k}) \leqslant \varepsilon_F \leqslant \varepsilon_{+}(\vec{k})$
	(`Pauli blocking'), and the
	conservation of energy $\varepsilon_+(k_x,k_y)-\varepsilon_-(k_x,k_y)=\hbar\omega$, for a given Fermi energy $\varepsilon_F$ and exciting frequency $\omega$. This means that only those points $(k_x,k_y)$ lying on the resonance curve $C_r(\omega)=\{(k_x,k_y)|
	2d(k_x,k_y)=\hbar\omega\}$ which satisfy the
	inequality $k_F^+(\theta)<k<k_F^-(\theta)$ will
	contribute to the joint density of states (JDOS) [see Fig.\,\ref{fig:LEVELSANDCONTOURS}(a)]. Here $k_F^{\lambda}(\theta)$ are the Fermi contours,
	defined by the equation $\varepsilon_{\lambda}(k_x,k_y)=\varepsilon_F$ which, written in polar coordinates, are given by
	$k^{\lambda}_{F}(\theta) = \sqrt{2m \varepsilon_{F}/\hbar^2+ k^{2}_{so}(\theta)} - \lambda k_{so}(\theta)$, where
	$ k_{so}(\theta) = mg_{[hkl]}(\theta)/\hbar^2 $ is a characteristic wave number of the SO interaction. The function $g_{[hkl]}(\theta)$ accounts for the anisotropy of the energy splitting $2d(k_x,k_y)=2kg_{[hkl]}(\theta)$, with
	$g_{[hkl]}(\theta)=|\boldsymbol{\mu}_x\cos\theta+\boldsymbol{\mu}_y\sin\theta|$, where the vectors $\boldsymbol{\mu}_\nu$, with components
	$(\boldsymbol{\mu}_{\nu})_{i} = \mu_{i\nu}$, are
	the columns of the matrix $\mu_{i\nu}$. The above restrictions imply that the allowed transitions are
	possible only in the energy window $\hbar\omega_+(\theta)\leqslant\hbar\omega\leqslant\hbar\omega_-(\theta)$, where $\hbar\omega_{\lambda}(\theta)=2d(k_F^{\lambda}(\theta))=2k_F^{\lambda}(\theta)g_{[hkl]}(\theta)$ is 
	the minimum (maximum) photon energy $\hbar\omega_+(\theta)$
	($\hbar\omega_-(\theta)$) required to induce a vertical transition between states lying along the direction $\theta$ in the $k$-space [inset in Fig.\,\ref{fig:LEVELSANDCONTOURS}(a)]. As a consequence, the JDOS for the system with SOI R+D$[hkl]$ reads as
	\begin{equation}
		J_{+-}(\omega)=\frac{\hbar\omega}{16\pi^{2}} \int \!\! d\theta\,\,\frac{\Theta[1-|\eta(\omega,\theta)|]}{g^{2}_{[hkl]}(\theta)}\ ,
	\end{equation}
	where $\Theta(x)$ is the Heaviside unit step function, and
	$\eta(\omega,\theta)=[\omega-\frac{1}{2}(\omega_-(\theta)+\omega_+(\theta))]/[\frac{1}{2}(\omega_-(\theta)-\omega_+(\theta))]$.
	According to this expression the edges of the absorption spectrum will be at the photon energies $\hbar\omega=\text{min}_{\theta}\{\hbar\omega_+(\theta)\}$ and $\hbar\omega=\text{max}_{\theta}\{\hbar\omega_-(\theta)\}$, which will occur along some directions $\theta_<$ and $\theta_>$ where the energy splitting is minimum and maximum, respectively. In general, the set of van Hove singularities can be identified geometrically by analyzing how the resonance curve $C_r(\omega)$ enters, intersects, and leaves the region of allowed transitions $k_F^+(\theta)<k<k_F^-(\theta)$ [shaded area in Fig.\,\ref{fig:LEVELSANDCONTOURS}(a)] as frequency varies. The resonance curve is a rotated ellipse with equation $|k_x\boldsymbol{\mu}_x+k_y\boldsymbol{\mu}_y|=\hbar\omega/2$. After a rotation by an angle $\zeta$ defined by $\tan 2\zeta=2\boldsymbol{\mu}_x\cdot\boldsymbol{\mu}_y/[|\boldsymbol{\mu}_x|^2-|\boldsymbol{\mu}_y|^2]$, the ellipse becomes of the standard form with principal axes
	given by $Q_x(\omega)=\hbar\omega/2g_{[hkl]}(\theta_<)$ and
	$Q_y(\omega)=\hbar\omega/2g_{[hkl]}(\theta_>)$, where $\theta_>=\theta_<+\pi/2$. In order to maintain the angle $\theta_<$ as the direction of global minimum energy separation (as defined above) we choose $\theta_<=\zeta$ if
	$g_{[hkl]}(\zeta) < g_{[hkl]}(\zeta + \pi/2)$,
	and $\theta_<=\zeta + \pi/2$ otherwise. The critical
	van Hove energies are given by those values of $\hbar\omega$ at which the axes $Q_x(\omega)$ and $Q_y(\omega)$ 
	of the resonance ellipse touch tangentially the Fermi contours
	$k_F^{\pm}(\theta)$ along $\theta_<$ and $\theta_>$.
	Figure\,\ref{fig:LEVELSANDCONTOURS}(a) illustrates the
	situation for the R+D$[123]$ case.
	The critical frequencies are then defined by the matchings $Q_x(\omega^+)=k_F^+(\theta_<)$,
	$Q_x(\omega_a)=k_F^-(\theta_<)$, $Q_y(\omega_b)=k_F^+(\theta_>)$, and
	$Q_y(\omega^-)=k_F^-(\theta_<)$:
	\begin{align}
		\hbar\omega^+ &=2 k^{+}_{F}(\theta_{<}) g_{[hkl]}(\theta_{<}), \label{OMEGA+} \\
		\hbar\omega_a & =2 k^{-}_{F}(\theta_{<}) g_{[hkl]}(\theta_{<}), \label{OMEGA<} \\
		\hbar\omega_b &= 2 k^{+}_{F}(\theta_{>}) g_{[hkl]}(\theta_{>}), \label{OMEGA>} \\
		\hbar\omega^- &= 2 k^{-}_{F}(\theta_{>}) g_{[hkl]}(\theta_{>}). \label{OMEGA-}
	\end{align}
	Note that while $\omega^+$ is always smaller than $\omega^-$, the order relation  between $\omega_a$ and $\omega_b$ can change for different orientations. 
	
	In Fig.\,\ref{fig:LEVELSANDCONTOURS}(b) we show the JDOS for
 different crystal orientations as a function of frequency, 
 keeping the same SO parameters values. The overall shape and size of the spectra reveals a strong dependence on the direction of sample growth. 
	For the case R+D$[111]$, the Hamiltonian is formally identical to that of a system with Rashba coupling only, 
	which presents an isotropic splitting of the states. Thus, the resonance curve and Fermi contours are concentric circles 
	and the JDOS displays the well known box-like shape with only two spectral features
	($\hbar\omega^{\pm}$). However, for other orientations the $k$-space for allowed optical transitions
	becomes no longer isotropic, and two types of shapes may appear, depending on the relative values of
	$ \omega_a $ and $ \omega_b $. When $\omega_a < \omega_{b}$ the spectra can develop a convex shape between these
	critical energies, as shown in Fig.\,\ref{fig:LEVELSANDCONTOURS}(b) for the R+D$[001]$ and R+$[123]$ systems.
	On the other hand, when $\omega_a > \omega_b$ the JDOS presents a linear dependence instead, as is illustrated by the
	R+D$[110]$ case. This can be explained by observing how the resonance curve $C_r(\omega)$ overlaps the region
	of allowed transitions bounded by the Fermi contours in each case. When $\omega_{a}<\omega_{b}$, the semi-axis $Q_{x}(\omega)$ 
	of the curve $C_r(\omega)$ will reach the line $k^{-}_{F}(\theta)$ first before the semi-axis $Q_{y}(\omega)$ 
	intersects the line $ k^{+}_{F}(\theta)$, which means that for $ \omega_{a} < \omega < \omega_b $ there is a portion 
	of the curve which does not contribute to the JDOS. 
	In contrast, when $ \omega_a > \omega_b $ we have the opposite situation, the semi-axis $Q_{x}(\omega)$ touch the 
	Fermi contour $ k^{-}_{F}(\theta) $ after the axis $ Q_{y}(\omega) $ contacts the Fermi contour $ k^{+}_{F}(\theta) $.
	This imply that there is a range of frequencies, $\omega_b < \omega < \omega_a $, for which the ellipses $ C_r(\omega) $ 
	lie entirely within the allowed zone (shaded area in Fig.\,\ref{fig:LEVELSANDCONTOURS}(a)), causing a linear increase in the JDOS. 
	The dependence of the critical frequencies (\ref{OMEGA+})-(\ref{OMEGA-}) on the
	Miller indices, suggests the growth direction as an additional element of control of the spectrum of optical transitions.
	
\subsection{First order spin current conductivity}
	
	The spin conductivity tensor at the fundamental frequency for the SO coupled 2DEG, as obtained from Kubo formula (\ref{SPINCONDLINEAR}), is 
	(no sum over repeated indices is implied)
	\begin{eqnarray}
	\operatorname{Re}\sigma^{\ell}_{ij}(\omega) &=& \sigma^{\ell}_{ij}(0) - \frac{e}{8\pi} \left( \boldsymbol{\mu_{x}} \times \boldsymbol{\mu_{y}} \right)_{\ell}  \frac{\hbar\omega}{8m/\hbar^2} \frac{1}{2\pi} \int^{2\pi}_{0}  \frac{d\theta}{g^4(\theta)} \ln \left| \frac{\left[ \omega + \omega_{+}(\theta) \right]\left[ \omega - \omega_{-}(\theta) \right]}{\left[ \omega - \omega_{+}(\theta) \right]\left[ \omega + \omega_{-}(\theta) \right]}\right| \label{Resigmal} \\
	&& \hspace*{8cm}		\times  \left[ \epsilon_{ijz} + \sin2\theta \left( \delta_{iy} - \delta_{ix} \right)\delta_{ij} + \cos2\theta \left( 1-\delta_{ij} \right) \right] \nonumber \\
	\operatorname{Im}\sigma^{\ell}_{ij}(\omega) &=& -\frac{e}{8\pi} \left( \boldsymbol{\mu_{x}} \times \boldsymbol{\mu_{y}} \right)_{\ell} \frac{\hbar\omega}{16m/\hbar^2} \int^{2\pi}_{0} \! \frac{d\theta}{g^4(\theta)} \left[ \epsilon_{ijz} + \sin2\theta \left( \delta_{iy} - \delta_{ix} \right)\delta_{ij} + \cos2\theta \left( 1-\delta_{ij} \right) \right] \Theta [ 1-|\eta(\omega,\theta)|],
	\label{Imsigmal}
	\end{eqnarray}
	 where
	\begin{equation}
	\sigma^{\ell}_{ij}(0) = - \frac{e}{8\pi} \left( \boldsymbol{\mu_{x}} \times \boldsymbol{\mu_{y}} \right)_{\ell} \frac{1}{2\pi} \int^{2\pi}_{0} \!  \frac{d\theta}{g^2(\theta)} \left[ \varepsilon_{ijz} + \sin2\theta \left( \delta_{iy} - \delta_{ix} \right)\delta_{ij} + \cos2\theta \left( 1-\delta_{ij} \right) \right],  
	\end{equation}
	is the dc spin conductivity. Note that $\sigma^{\ell}_{yy}(\omega)=-\sigma^{\ell}_{xx}(\omega)$.
	Complex integration gives the result 
	\begin{equation} \label{dc}
	\sigma^{\ell}_{ij}(0)=-\frac{e}{8\pi}\frac{(\boldsymbol{\mu}_x\times \boldsymbol{\mu}_y)_{\ell}}{|\boldsymbol{\mu}_x\times\boldsymbol{\mu}_y|}
\left(\begin{array}{cc}
-\mbox{Im}\,(z_+) & 1+\mbox{Re}\,(z_+) \\
\\
-1+\mbox{Re}\,(z_+) & \mbox{Im}\,(z_+)
\end{array}\right) \ ,
	\end{equation}
	where 
\begin{equation}
z_+=-\frac{[A-\sqrt{A^2-(B^2+C^2)}]}{B^2+C^2}(B+iC) \ ,
\end{equation}
with $A=(|\boldsymbol{\mu}_x|^2+|\boldsymbol{\mu}_y|^2)/2$,
$B=(|\boldsymbol{\mu}_x|^2-|\boldsymbol{\mu}_y|^2)/2$, and
$C=\boldsymbol{\mu}_x\cdot\boldsymbol{\mu}_y$. 
The general expression (\ref{dc}) extends the result reported in Ref.\,\cite{Chen2014},
which is valid for SO vector fields with $d_z({\bf k})=0$ ($\mu_{zx}=\mu_{zy}=0$) only. 
	
\begin{figure}
\centering
\includegraphics[scale=0.35]{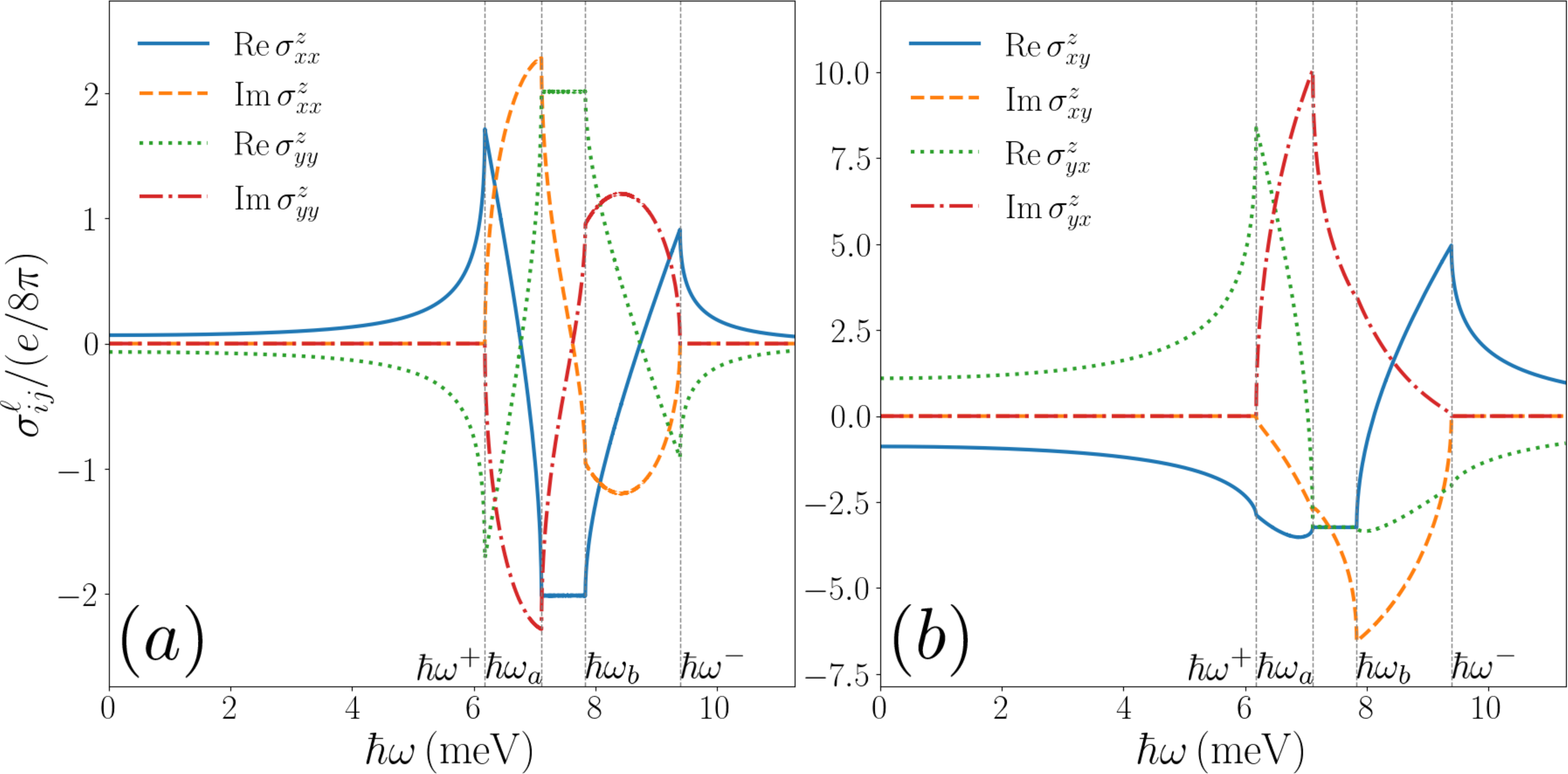}
\caption{Longitudinal (a) and transverse (b) components of the linear spin current conductivity $\sigma^z_{ij}(\omega)$ for a R+D[$ 123 $] 2DEG. The vertical dotted lines indicate the positions of the critical frequencies.
The parameters used are the same as in Fig.\,\ref{fig:LEVELSANDCONTOURS}.}
\label{fig:SPINLINEALMUIJKJ}
\end{figure}
Since $\left( \boldsymbol{\mu_{x}} \times \boldsymbol{\mu_{y}} \right)_{x}$ and $\left( \boldsymbol{\mu_{x}} \times \boldsymbol{\mu_{y}} \right)_{y}$ vanish for $ \mu_{z\nu} = 0 $ we have that the only systems supporting a linear spin current with perpendicular-to-plane spin polarization strictly, are those grown along the $ [001] $ and $ [111] $ directions; any other crystal orientation will have in-plane spin polarized current components. Moreover, the vanishing of the common factor $\boldsymbol{\mu_{x}} \times \boldsymbol{\mu_{y}}$ implies the
	absence of an induced spin current via electric-dipole interaction in the 2DEG with R+D$[hkl]$ SO coupling. This reminds
	the well known effects due to the recovery of the SU(2) symmetry predicted in systems with R+D$[001]$, like the infinite spin lifetime due
	to fixed spin precession axis \cite{Schliemann2017} or the formation of a persistent spin helix state \cite{Bernevig2006,Dettwiler2017}.

	As we mentioned before, Kammermeier et al.\cite{Kammermeier2016a} found that for an arbitrary crystal orientation is still possible to have conditions for spin-preserving symmetries due to the interplay of Rashba and Dresselhaus SOI. 
	The requirement for that is to have samples with two Miller indices equal in modulus and a particular relation between the Rashba and Dresselhaus 
	parameters (in our language, a proper combination of the elements of the SO matrix $\mu_{ij}$) \cite{Kammermeier2016a}.
	It can be verified that for these special conditions, the factor $\boldsymbol{\mu_{x}} \times \boldsymbol{\mu_{y}}$ is zero. 
	Without loss of generality, lets choose, after Kammermeier, the orientation $\vec{\hat{n}} = (\eta,\eta,n_z)$, with $ \vec{\hat{m}}=(-1,1,0)/\sqrt{2} $ and $ \vec{\hat{l}} = (n_z, n_z, -2\eta)/\sqrt{2} $, where $n_z^2=(1-2\eta^2)$. The
	corresponding vectors of SO parameters become $\boldsymbol{\mu_x} = (0, -\alpha + \gamma k^{2}_{n}(1-9\eta^2)n_z, 0) $ and $ \boldsymbol{\mu_y} = (\alpha + \gamma k^{2}_{n}(1+3\eta^2)n_z, 0, -\gamma k^{2}_{n} \sqrt{2}\eta (1-3\eta^2))$, where
	$\alpha$ and $\gamma k_n^2$ are the Rashba and Dresselhaus coupling strengths, respectively. When these satisfy the relation $\alpha/\gamma k_n^2=(1-9\eta^2)n_z$
	($\mu_{yx}=0$), the SO vector field becomes collinear,
	${\bf d}({\bf k})=\gamma k_n^2(3\eta^2-1)(-2n_z,0,\sqrt{2}\,\eta)k_y$.
Under these conditions, $\boldsymbol{\mu_{x}} \times \boldsymbol{\mu_{y}}=\mu_{yx}(\mu_{zy},0,-\mu_{xy})=0$, implying the vanishing of the spin current. Note also that when $\mu_{yx}\neq 0$, and $\mu_{zy}=0$ (which is true for the $[111]$ growth direction only, $\eta^2=1/3$) or $\mu_{xy}=0$, the polarization of the spin current is along the direction ${\bf\hat{n}}$ ($z'$) or ${\bf\hat{l}}$ ($x'$), respectively. In contrast, the spin current polarized parallel to the ${\bf\hat{m}}$ direction is null regardless the magnitude of the SO strength 
	parameters, $\sigma^y_{ij}(\omega)=0$.
	Interestingly, although the condition $\mu_{xy}=0$, which rewrites as
	$\alpha/\gamma k^{2}_{n} = -(1+3\eta^2)n_z$,
	does not corresponds to a collinear SO vector field, still produces an
	absence of out-of-plane-polarized spin current.
	
	In Fig.\,\ref*{fig:SPINLINEALMUIJKJ}, a typical spectrum of the linear
	spin current conductivity $\sigma^z_{ij}(\omega)$ is displayed, in this case for the R+D$[123]$ system. 
	As anticipated by the JDOS, we can identify the presence of van Hove features
	at the critical energies (\ref{OMEGA+})-(\ref{OMEGA-}),
	defined by the Pauli blocking and the energy conservation condition for vertical
	transitions (Fig.\,\ref{fig:LEVELSANDCONTOURS}(b)). 
	Given that the critical frequencies (\ref{OMEGA+})-(\ref{OMEGA-}) depend not only on the magnitude and relative value of the SO material parameters but also on the crystal orientation, our results
	suggests a kind of spectral control of the overall shape of the linear spin current response by choosing appropriately the samples in advance. 
	
	\subsection{SH spin current conductivity}

	The second-order Kubo formula (\ref{2OMEGATWOLEVEL}) leads to the SH spin conductivity of the R+D$[hkl]$ system,
	\begin{eqnarray}
	\sigma^{\ell,(2\omega)}_{ijl}(\omega)&=&\frac{e^2}{(\hbar\omega)^2} \frac{(\hbar^2/m)}{(2\pi)^2}
	(\boldsymbol{\mu_{x}} \times \boldsymbol{\mu_{y}})_p\int^{2\pi}_{0} \! \frac{d\theta}{g^{6}(\theta)} \hat{k}_i\hat{k}_{\nu}
	\left\{ (\boldsymbol{\mu_{x}} \times \boldsymbol{\mu_{y}})_p\mu_{\ell\nu}(\delta_{jl}-\hat{k}_j\hat{k}_l)
	C(\omega,\theta) \right. \\
	&& \hspace*{5.5cm} \left. -\epsilon_{\ell p q}\mu_{q\rho}\hat{k}_{\rho}(\boldsymbol{\mu_{j}} \cdot \boldsymbol{\mu_{\nu}})
	(\hat{k}_x\delta_{ly}-\hat{k}_y\delta_{lx}) [C(\omega,\theta)-4C(2\omega,\theta)]+ (j\leftrightarrow l)\right\} \nonumber
	\end{eqnarray}
	where $ \hat{k}_{i}(\theta) = \cos\theta \delta_{ix} + \sin\theta \delta_{iy} $ and
	\begin{equation}
	 C(x,\theta)=-\frac{1}{4} \left( \frac{2mg^2(\theta)}{\hbar^2} + \frac{\hbar x}{4} \operatorname{ln} \bigg\vert \frac{[x+\omega_{+}(\theta)][x-\omega_{-}(\theta)]}{[x-\omega_{+}(\theta)][x+\omega_{-}(\theta)]} \bigg\vert \right)
	 -i\pi\frac{\hbar x}{16}\,\Theta[1-|\eta(x,\theta)|].
	\end{equation}
	
	As in the linear response, the system grown along $\vec{\hat{n}}=(\eta,\eta,n_z)$
	will not support a spin current at $2\omega$ when $\mu_{yx}=0$, or equivalently when the
	SO field $d_{i}({\bf k})=\mu_{i\nu}k_{\nu}$ becomes collinear.
	However, for the same special class of orientations, to analyze the spin conductivity
	for each spin index $\ell$, we have to focus in the following factors
	\begin{align}
	 	(\boldsymbol{\mu_{x}})_{\ell} & = \mu_{yx}\delta_{\ell y} , \label{SHFACTOR1}\\
		(\boldsymbol{\mu_{y}})_{\ell} & = \left( \mu_{xy}\delta_{\ell x} + \mu_{zy} \delta_{\ell z} \right) ,\label{SHFACTOR2}\\
		\big[ \left( \boldsymbol{\mu_{x}} \times \boldsymbol{\mu_{y}} \right) \times \boldsymbol{\mu_{x}} \big]_{\ell} & = \mu^{2}_{yx} \big( \mu_{xy}\ \delta_{\ell x} + \mu_{zy}\ \delta_{\ell z} \big) ,\label{SHFACTOR3}\\
		\big[ \left( \boldsymbol{\mu_{x}} \times \boldsymbol{\mu_{y}} \right) \times \boldsymbol{\mu_{y}} \big]_{\ell}  & = -\mu_{yx} \left( \mu^{2}_{xy} + \mu^{2}_{zy} \right) \delta_{\ell y} \label{SHFACTOR4}.		
	\end{align}

\begin{figure}
\centering
\includegraphics[scale=0.35]{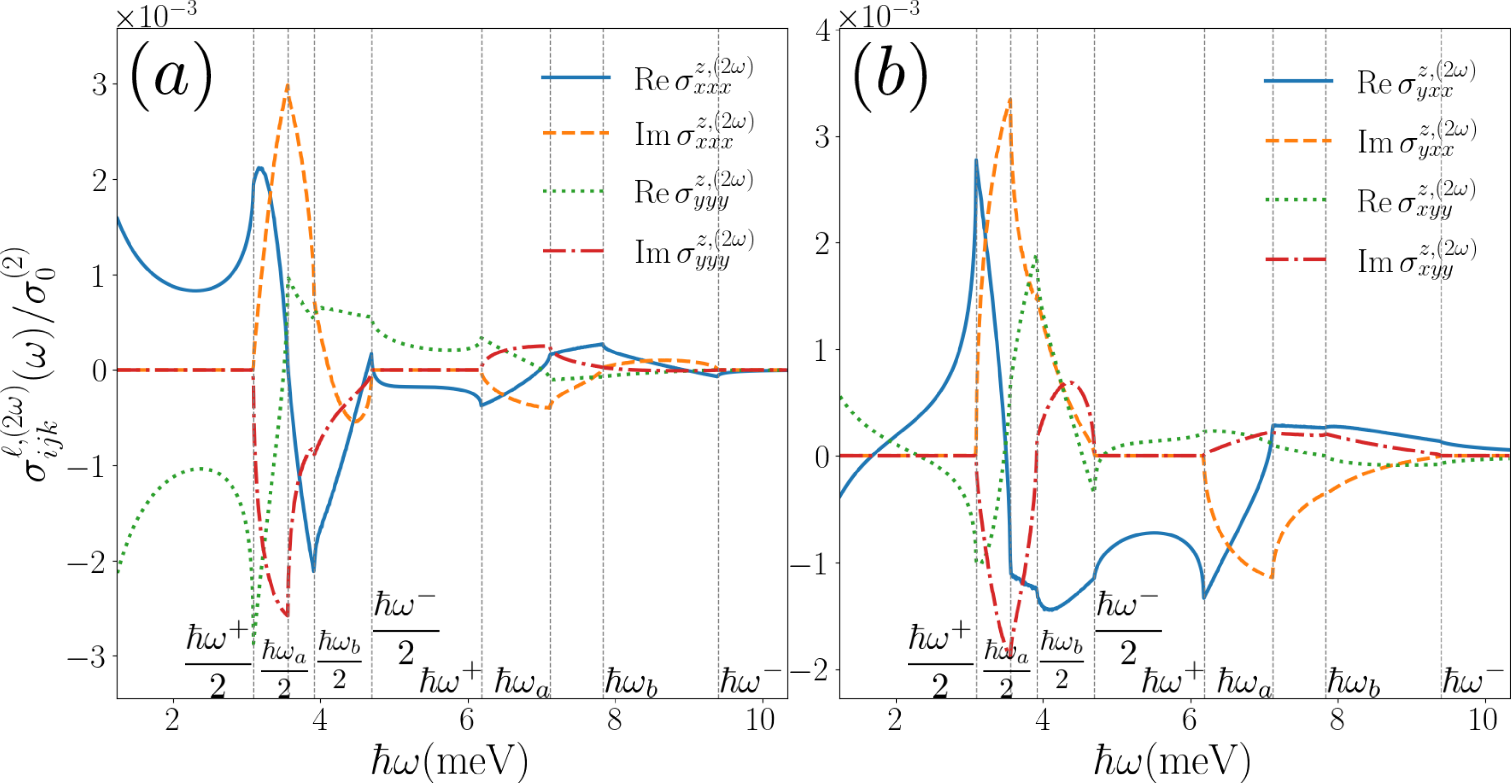}
\caption{Longitudinal (a) and Hall (b) components of the second-harmonic spin current conductivity tensor with out-of plane spin orientation for a R+D[$ 123 $] 2DEG, where $ \sigma^{(2)}_{0} = e^2 \alpha/4\pi\varepsilon^{2}_{R} $, $\,\varepsilon_{R} = m \alpha^2/ \hbar^2$. The parameters used are the same as in Fig.\,\ref{fig:LEVELSANDCONTOURS}.}
\label{fig:SPIN2OMEGAMUIJKJ}
\end{figure}
	
	We note that there are some possibilities to control the polarization of the nonlinear spin current.
	For $\mu_{yx}\neq 0$, the condition $\mu_{xy}=0$ ($\mu_{zy}=0$) corresponds to have
	a spin current flowing in the plane defined by ${\bf\hat{m}}$ and ${\bf\hat{n}}$
	(${\bf\hat{l}}$ and ${\bf\hat{m}}$), that is $\sigma^{x,(2\omega)}_{ijl}(\omega)=0$ ($\sigma^{z,(2\omega)}_{ijl}(\omega)=0$). On the other hand, to have $\sigma^{y,(2\omega)}_{ijl}(\omega)=0$ is possible only for $\mu_{yx}=0$. Therefore, we have that for a system with crystal orientation $\vec{\hat{n}} = (\eta,\eta,n_z)$, and given that
	the linear $\sigma^y_{ij}(\omega)=0$, the generation of a spin current
	polarized along the $ \vec{\hat{m}} $ ($ y' $) direction will depend quadratically on the electric field, because it is induced as
	a second-order response strictly.

	Fig.\,\ref*{fig:SPIN2OMEGAMUIJKJ} shows the SH spin current conductivity
	$\sigma^{z,(2\omega)}_{ijl}(\omega)$ for the R+D$[123]$ system.
	As expected, beside the peaks related with critical points at
	the energies $\hbar\omega^{\pm},\hbar\omega_a,\hbar\omega_b$, van Hove 
	singularities appear also at their subharmonics. Similarly to the first-order
	spin current conductivity, the magnitude and direction of the nonlinear spin current
    could be modified through frequency variation, relative value of the SO strengths,
    or by a proper choice of the sample growth direction.
    
    Another aspect worth noting is that the R+D$ [hkl] $ systems will present a nonlinear spin Hall effect, through $ \sigma^{z,(2\omega)}_{yxx}(\omega) \neq 0 $
    and $ \sigma^{z,(2\omega)}_{xyy}(\omega) \neq 0 $, except for R+D$ [001] $ and R+D$ [111] $ cases. The necessary condition for this phenomenon is to have $ d_z(\vec{k}) \neq 0 $ $ (\mu_{z\nu} \neq 0) $, which is characteristic of $ [110] $ samples, one of the low Miller indices usually studied.

	\section{2D anisotropic Rashba model} \label{ARM}
	
	\subsection{Energy spectrum and the joint density of states}
	
	In this section we considered another anisotropic model, 
	used to study the effect of an anisotropic Rashba splitting on the longitudinal
	optical conductivity of 2D puckered structures \cite{Saberi-Pouya2017} like black phosphorus and group IV monochalcogenides \cite{Gomes2015,Kamal2016,Zeraati2016}.
	Here the reduced symmetry of the splitting of the states is due to mass anisotropy.
	The kinetic contribution to the low energy Hamiltonian is that of an anisotropic free electron gas \cite{Ahn2021,Ahn2021a},
	$ \varepsilon_{0}(\vec{k}) = \hbar^2 k^{2}_{x}/2m_x + \hbar^2 k^{2}_{y}/2m_y $,
	while the Rashba SO field is taken as $ \vec{d}(\vec{k}) = \alpha ( \sqrt{m_d/m_y} k_y \vec{\hat{x}} - \sqrt{m_d/m_x} k_x \vec{\hat{y}} ) + \Delta \vec{\hat{z}} $,
	where $\alpha$ is the SO strength and $m_d=\sqrt{m_xm_y}$ is the geometric mean
	of the masses $m_x$ and $m_y$ along the $x$- and $y$- directions \cite{Popovic2015}. The model has been extended to include an energy parameter, $\Delta\geqslant 0$, which gives rise to a gap in the energy dispersion; by taking
 $\Delta=0$ or $\Delta\neq 0$ we can move from a model with TRS to a model with broken TRS.
 This massive anisotropic
	Rashba low-energy model could describe a gapped conduction band of the phosphorene
	monolayer \cite{Saberi-Pouya2018}.
	
	Written in polar coordinates, the conduction ($\lambda=+$) and valence ($\lambda=-$)  bands are $\varepsilon_{\lambda}(k,\theta) = \hbar^2 k^2 g^2(\theta)/2m_d + \lambda \sqrt{\alpha^2 k^2 g^2(\theta) + \Delta^2} $. The function $g(\theta) = [(m_d/m_x) \cos^2\theta + (m_d/m_y) \sin^2\theta]^{1/2}$ measures the separation 
	of energy-constant curves along the direction $\theta$ in the $k$-space; 
	note that $g(\theta)=1$ when $ m_x = m_y$, which corresponds to the well known
	case of a magnetized 2DEG with Rashba coupling in quantum wells of semiconductor heterostructures \cite{Dyrda2017}. 
	The energy difference between the bands is $ \varepsilon_{+}(\vec{k}) - \varepsilon_{-}(\vec{k}) = 2d(\vec{k}) = 2 \sqrt{\alpha^2 k^2 g^2(\theta)+\Delta^2}  $ and the constant energy-difference curve, $ C_r (\omega) = \{ (k_x,k_y) \vert \varepsilon_{+}(\vec{k}) - \varepsilon_{-}(\vec{k}) = \hbar\omega \} $ 
	is the ellipse with equation $ \varepsilon_{0}(k_x,k_y) = [(\hbar\omega/2)^2 - \Delta^2] / 2\varepsilon_R $, where $ \varepsilon_R = m_d\alpha^2 / \hbar^2 $
	is a characteristic energy associated to the Rashba interaction. 	
	\begin{figure}
		\centering
		\includegraphics[scale=0.25]{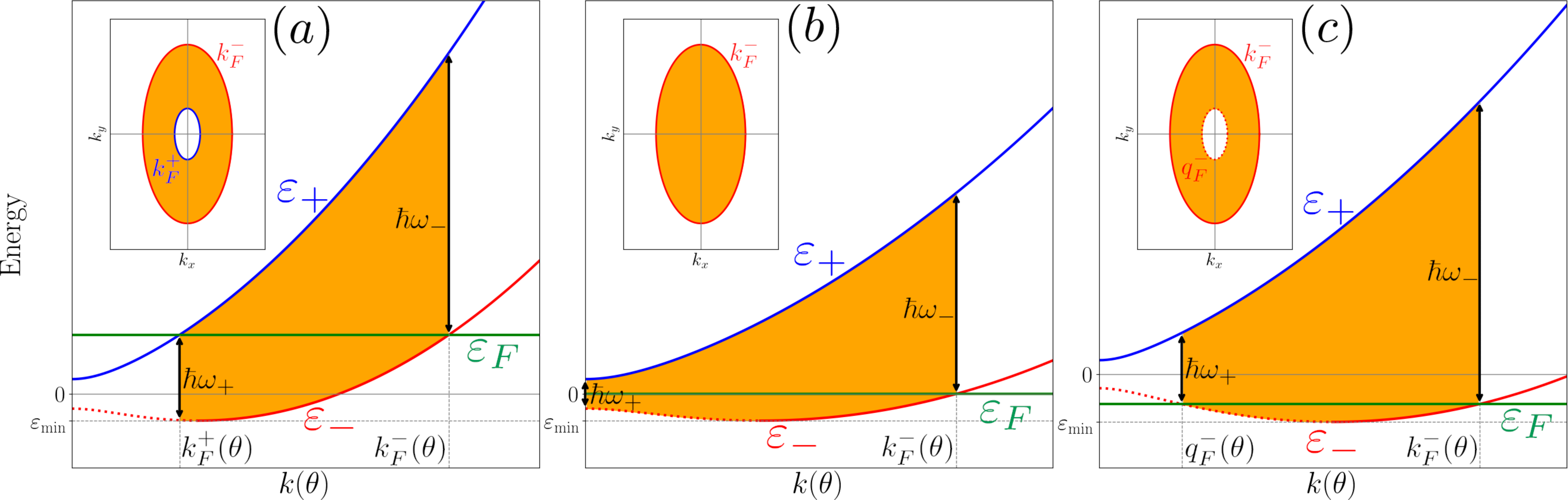}
		\caption{Energy bands $\varepsilon_{\pm}(k,\theta)$
		of a gapped 2D anisotropic Rashba model when $\varepsilon_R>\Delta$. 
		The shaded areas indicate the $k$-region of allowed optical transitions
		for several Fermi level positions: (a) $ \varepsilon_{F} > \Delta $, (b) $ \vert\varepsilon_F\vert < \Delta $ and (c) $ \varepsilon_{min} < \varepsilon_F < -\Delta $. The insets show the respective Fermi contours.}
		\label{NIVELESCONTORNOSANIS}
	\end{figure}
	
	The shape of the band $\varepsilon_{-}(\vec{k})$ depends on the ratio $p=\Delta/\varepsilon_R$. When $p<1$, the band acquires a mexican hat shape,
	with a local maximum $-\Delta$ at the origin ${\bf k}={\bf 0}$ and two local minimums
	of value $ \varepsilon_{min} = -(\varepsilon^{2}_{R}+\Delta^2)/2\varepsilon_{R} $
	at ${\bf k}$-points lying on the ellipse $\varepsilon_{0}(k_x,k_y) = 
	(\varepsilon^{2}_{R} - \Delta^2)/2\varepsilon_{R}$. Otherwise, the valence band develops only a minimum at the origin.
	This means that there are several distinct positions for the Fermi level: $(i)\,
	\varepsilon_{F}>\Delta$ (Fig.\,\ref{NIVELESCONTORNOSANIS}(a)), and then two Fermi contours are generated
	\begin{equation}\label{KF+-}
		k^{\pm}_{F}(\theta) = \frac{1}{\alpha g(\theta)} \left[ \left( \sqrt{\varepsilon^{2}_{R} +\Delta^2 + 2\varepsilon_{R}\varepsilon_F } \mp \varepsilon_{R} \right)^2 - \Delta^2 \right]^{1/2},
	\end{equation}
	from equations $\varepsilon_+(k,\theta)=\varepsilon_-(k,\theta)=\varepsilon_F$;
	$(ii)\, |\varepsilon_F|<\Delta $ (Fig.\,\ref{NIVELESCONTORNOSANIS}(b)), 
	where there is only one Fermi contour
	$ k^{-}_{F}(\theta) $, lying on the valence band; and $(iii)$ if $p<1$, 
	$\varepsilon_{min} < \varepsilon_{F} < -\Delta$ 
	(Fig.\,\ref{NIVELESCONTORNOSANIS}(c)), where the Fermi lines arise only 
	from the valence band through the equation $\varepsilon_-(k,\theta)=\varepsilon_F$,
	with roots $ k^{-}_{F}(\theta) $ and
	\begin{equation}\label{QF-}
		q^{-}_{F}(\theta) = \frac{1}{\alpha g(\theta)} \left[\left( \varepsilon_{R} - \sqrt{\varepsilon^{2}_{R} +\Delta^2 + 2\varepsilon_{R}\varepsilon_F }\right)^2 - \Delta^2 \right]^{1/2}.
	\end{equation}
	Note that the situation $(i)$ or $(iii)$ includes the gappless case $ \Delta = 0 $, the Fermi level being then positive or negative, respectively. 
	
	Interestingly, the Fermi lines described by (\ref{KF+-}) and (\ref{QF-}) are concentric ellipses vertically (horizontally) oriented if $ m_y > m_x $ ($ m_x > m_y $), see insets in Fig.\,\ref{NIVELESCONTORNOSANIS}. The energy separation of the bands at these lines is independent of the direction $ \theta $ in $ k $-space, taking the values
	$2d(k^{\pm}_{F}(\theta)) = 2\left( \sqrt{\varepsilon^{2}_{R} + \Delta^2 + 2\varepsilon_{R}\varepsilon_{F}}  \mp \varepsilon_{R} \right)$ or
	$2d(q^{-}_{F}(\theta)) = 2\left(\varepsilon_{R} - \sqrt{\varepsilon^{2}_{R} + \Delta^2 + 2\varepsilon_{R}\varepsilon_{F}} \right)$, according to 
	the position of the Fermi level.
\begin{figure}
\centering
\includegraphics[scale=0.25]{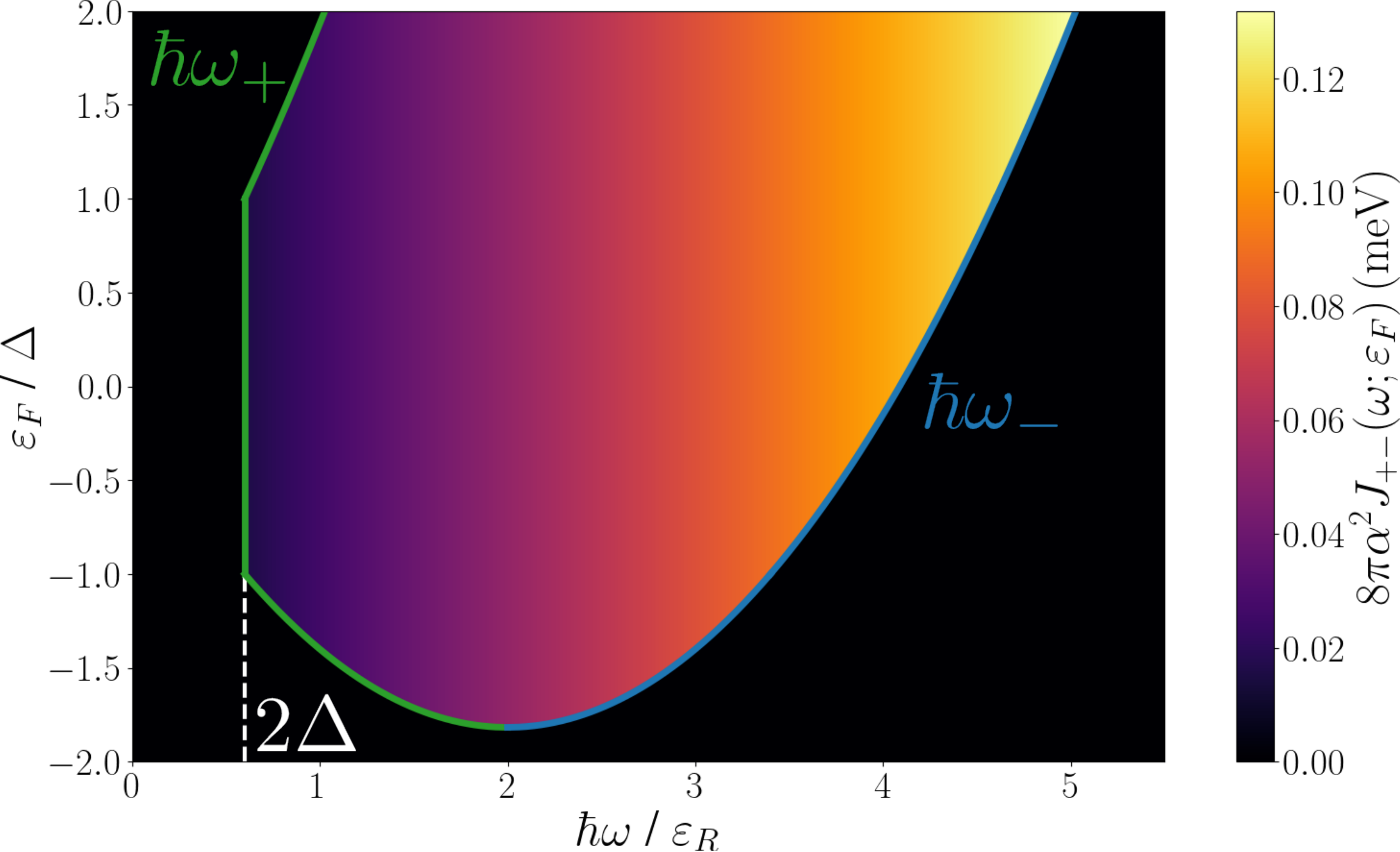}
\caption{Joint density of states $ 8\pi\alpha^2 J_{+-}(\omega;\varepsilon_{F}) $ for a gapped 2D anisotropic Rashba model with $ \varepsilon_{R} > \Delta$, 
and the absorption edges $\hbar\omega_{\pm}(\varepsilon_F)$.
The parameters used are $ \Delta = 0.3\, \varepsilon_R $, 
$ \alpha = 10\, \mathrm{meV} $\r{A}  , and 
$ m_{x}=m_0 $, $ m_{y}=4m_{0} $ for the effective masses. }
\label{fig:JDOSCOLORTICKS}
\end{figure}
	
	 The lowest (highest) energy of the spectrum of allowed interband transition will be denoted by $ \hbar\omega_+$ ($ \hbar\omega_- $), see Fig.\,\ref{NIVELESCONTORNOSANIS}. When $ \vert\varepsilon_{F}\vert < \Delta $ we have that the lowest possible transition occurs at the energy gap $ \hbar\omega_+ = 2\Delta$.  When $ \varepsilon_{F} > \Delta $ or $ \varepsilon_{min} < \varepsilon_{F} < -\Delta $, we have $ \hbar\omega_+ = 2d(k^{+}_{F}) $ or $ \hbar\omega_+ = 2d(q^{-}_{F}) $, respectively. The highest possible energy transition is always given by $ \hbar\omega_{-} = 2d(k^{-}_{F}) $.
	Moreover, the resonance curve $C_r(\omega)$ is an ellipse with the same shape and orientation than the Fermi lines, but differing only in size given its frequency dependence. As a consequence, there will be only two critical points in the JDOS,  
	given by the frequencies $\omega_+$ and $\omega_-$ at which the ellipse $C_r(\omega)$ enters and leaves, respectively, the region of allowed transitions (shaded areas in the
	insets of Fig.\,\ref{NIVELESCONTORNOSANIS}), and which  defines an absorption window $\hbar\omega_+(\varepsilon_F)<\hbar\omega<\hbar\omega_-(\varepsilon_F)$. All this is apparent in the JDOS
	\begin{equation}\label{key}
		J_{+-}(\omega;\varepsilon_F) = \frac{\hbar\omega}{8\pi\alpha^2} \Theta(1-\vert \eta(\omega,\varepsilon_F) \vert), \ \ \ \eta(\omega,\varepsilon_F) = \frac{\omega - (\omega_{-} + \omega_{+})/2}{ (\omega_{-} - \omega_{+})/2}\,,
	\end{equation}
	which is displayed as a color map in Fig.\,\ref{fig:JDOSCOLORTICKS}, for a system having $p<1$. For a given value of $\varepsilon_F$ the color gradation shows the
	linear dependence on the exciting frequency.
	
	\subsection{First order spin current conductivity}
	
	According to the formula (\ref{SPINCONDLINEAR}), the induced spin current
	response function of the anisotropic Rashba model becomes
	\begin{equation}\label{ELLij}
		\sigma^{\ell}_{ij}(\omega) = - \delta_{\ell z} \frac{e}{8\pi} \frac{1}{2\varepsilon_{R}} \left[ \varepsilon_{ijz} - i \frac{m_d}{m_i}\! \left(\frac{2\Delta}{\hbar\tilde{\omega}}\right) \delta_{ij} \right] \left[ A(\tilde{\omega}) + \frac{1}{2}\hbar(\omega_{-}-\omega_{+}) \right],
	\end{equation}
	 where
	\begin{equation}\label{AX}
		A(x) = \frac{\hbar x}{4}\left[1-\left(\frac{2\Delta}{\hbar x}\right)^2\right]
		\operatorname{ln}\left[\frac{(x+\omega_{+})(x-\omega_{-})}
		{(x-\omega_{+})(x+\omega_{-})}\right] \ .
	\end{equation}

\begin{figure}
\centering
\includegraphics[scale=0.32]{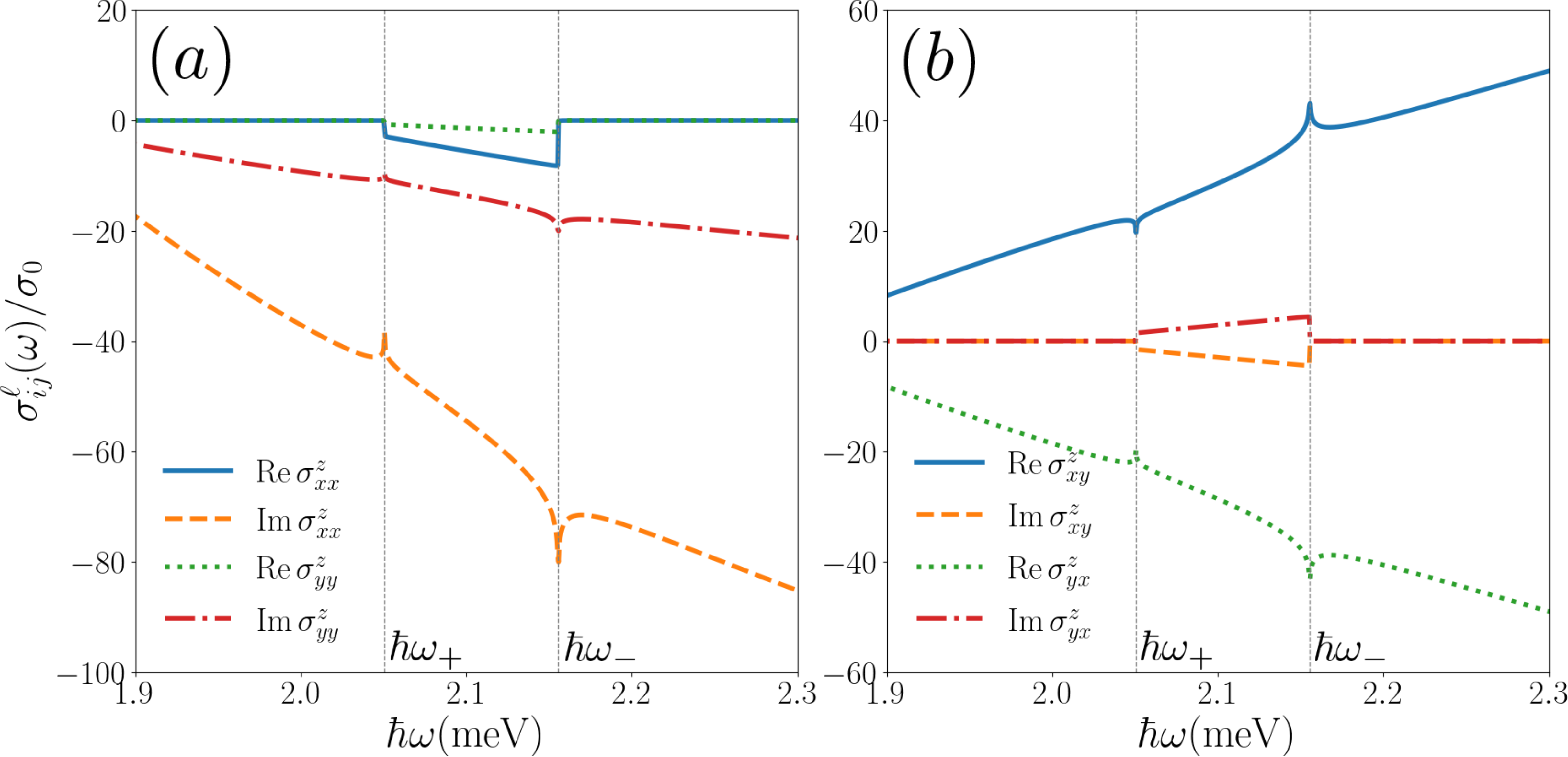}
\caption{Longitudinal (a) and Hall (b) components of the linear spin current conductivity tensor $\sigma^z_{ij}(\omega)$ for a gapped  anisotropic Rashba model, normalized to $ \sigma_0 = e/8\pi $. The parameters used are $ \Delta = 1\,\mathrm{meV} $, $ m_{x}=m_0 $, $ m_{y}=4m_{0} $, $ \alpha = 10 \,\mathrm{meV}$\r{A}, $ \varepsilon_{F} = 2\Delta $.}
\label{fig:SPINLINEAL2DELTA}
\end{figure}

 Remarkably, the linear spin currents generated in this system will have electrons with out-of-plane spin orientations only. This a consequence of the breaking of TRS by the term $d_z({\bf k})$, which is a non null constant in the present model. This makes the product of matrix elements $\langle \lambda | \mathcal{\hat{J}}^{\ell}_{i} | -\lambda \rangle \langle -\lambda | \hat{v}_j | \lambda \rangle$
 an odd (even) function in ${\bf k}$-space when $\ell=x,y$ ($\ell=z$).
Thus, the Kubo expression (\ref{SPINCONDLINEAR}) integrates to a non zero value when $\ell=z$ only. 
 The longitudinal components of the spin current conductivity (Fig.\,\ref{fig:SPINLINEAL2DELTA}(a)) are proportional to the gap parameter, $\sigma^{z}_{ii}(\omega)\propto\Delta$, and therefore they vanish
	for the gapless case. Moreover, these diagonal components are inversely proportional to $\sqrt{m_i}$, such that $m_x\sigma^z_{xx}(\omega)=
     m_y\sigma^z_{yy}(\omega)$, as a consequence of the
	mass anisotropy. On the other hand, the Hall components are nonzero, regardless of the value of $\Delta$, indicating the generation of a spin Hall effect in the gapped
	or ungapped system. In addition, $ \sigma^{\ell}_{xy}(\omega) = -\sigma^{\ell}_{yx}(\omega) $, as can be seen in Fig.\,\ref{fig:SPINLINEAL2DELTA}(b).
	The effect of the position of the Fermi level with respect to the gap, manifests
	through the critical energies $\hbar\omega_{\pm}(\varepsilon_F)$. The variation of Fermi energy leads mainly to a change of the window $\hbar(\omega_--\omega_+)$,
	just like in JDOS. 
	
	If ${\bf E}^{\omega}_0=E_0^{\omega}(\cos\varphi{\bf\hat{x}}+\sin\varphi{\bf\hat{y}})$ is the amplitude of the external field, the
	spin current can be written as the sum of a component along ${\bf E}^{\omega}_0$
	and a component perpendicular to it, 
	\begin{equation} \label{SCphi}
	{\boldsymbol{\cal{J}}^z}(\omega)=[\cos^2\varphi\,\sigma^z_{xx}(\omega)+\sin^2\varphi\,
	\sigma^z_{yy}(\omega)]{\bf E}^{\omega}_0+[\sin\varphi\cos\varphi
	(\sigma^z_{xx}(\omega)-\sigma^z_{yy}(\omega))+
	\sigma^z_{xy}(\omega)]({\bf E}^{\omega}_0\times {\bf\hat{z}})\ ,
	\end{equation}
	which reduces to ${\boldsymbol{\cal{J}}^z}(\omega)=\sigma^z_{xy}(\omega)({\bf E}^{\omega}_0\times {\bf\hat{z}})$ when $\Delta=0$. Expression
	(\ref{SCphi}) suggests how the induced spin current could be manipulated through
	the frequency dependence of the tensor $\sigma^z_{ij}$ and the direction of the
	applied in-plane electric field.
	
\subsection{SH spin current conductivity}
	
For the second-harmonic spin current conductivity we obtain the following expressions
from (\ref{2OMEGATWOLEVEL}) (no sum over repeated indices is implied),
\begin{gather}
\sigma^{\ell,(2\omega)}_{ijj}(\omega) = \delta_{\ell j} 	\varepsilon_{ijz} \frac{e^2 / 8\pi}{(\hbar\tilde{\omega})^2} \frac{1}{k_R} \left( \frac{m_d}{m_j} \right)^{1/2} F(\tilde{\omega}), \label{SHijj}\\
\sigma^{\ell,(2\omega)}_{iii}(\omega) = 	\frac{e^2/8\pi}{(\hbar\tilde{\omega})^2} \frac{1}{k_{R}} \left(\frac{m_{d}}{m_i}\right)^{3/2} \bigg\{\varepsilon_{ \ell iz}\ G(\tilde{\omega}) + 2i \left(1-\delta_{\ell z}\right) \left(\frac{2\Delta}{\hbar\tilde{\omega}}\right) \Big[A(\tilde{\omega}) - A(2\tilde{\omega})\Big] \bigg\},
\label{SHiii}\\
\sigma^{\ell,(2\omega)}_{iji}(\omega) = 	\sigma^{\ell,(2\omega)}_{iij}(\omega) = \frac{e^2/16\pi}{(\hbar\tilde{\omega})^2} \frac{1}{k_R} \left(\frac{m_d}{m_i}\right)^{1/2} \bigg\{\delta_{\ell i}\varepsilon_{ijz}\ H(\tilde{\omega}) + 2i (1-\delta_{\ell z})\left(\frac{2\Delta}{\hbar\tilde{\omega}}\right) \Big[A(\tilde{\omega}) - A(2\tilde{\omega})\Big]\bigg\},
\label{SHiji}
\end{gather}
where $ k_R =  \varepsilon_{R}/\alpha = m_d\alpha / \hbar^2 $, $ A(x) $ is given by (\ref{AX}), and
\begin{align}
F(\tilde{\omega}) &=  \Bigg\{ 1 + 	\frac{1}{4}\left[1-\left(\frac{2\Delta}{\hbar\tilde{\omega}}\right)^2 \right] \Bigg\} A(\tilde{\omega}) - \left[1-\left(\frac{2\Delta}{2\hbar\tilde{\omega}}\right)^2\right]A(2\tilde{\omega}) +\frac{1}{8}\hbar(\omega_{-}-\omega_{+}), \\
G(\tilde{\omega})&= \Bigg\{ 1-\frac{3}{4} \left[1- 	\left(\frac{2\Delta}{\hbar\tilde{\omega}} \right)^2  \right] \Bigg\} A(\tilde{\omega}) - \Bigg\{ 4 - 3\left[ 1-\left(\frac{2\Delta}{2\hbar\tilde{\omega}}\right)^2 \right] \Bigg\} A(2\tilde{\omega}) - \frac{3}{8}\hbar (\omega_{-} - \omega_{+}),\\
H(\tilde{\omega})&= \Bigg\{ 2 - \frac{1}{2}\left[ 1 - 	\left(\frac{2\Delta}{\hbar\tilde{\omega}}\right)^2 \right] \Bigg\} A(\tilde{\omega}) - 2 \Bigg\{ 2 - \left[ 1 - \left(\frac{2\Delta}{2\hbar\tilde{\omega}}\right)^2 \right] \Bigg\} A(2\tilde{\omega}) - \frac{1}{4}\hbar(\omega_{-} - \omega_{+}). \label{HDEOMEGA}
\end{align}
	
\begin{figure}
\centering
\includegraphics[scale=0.26]{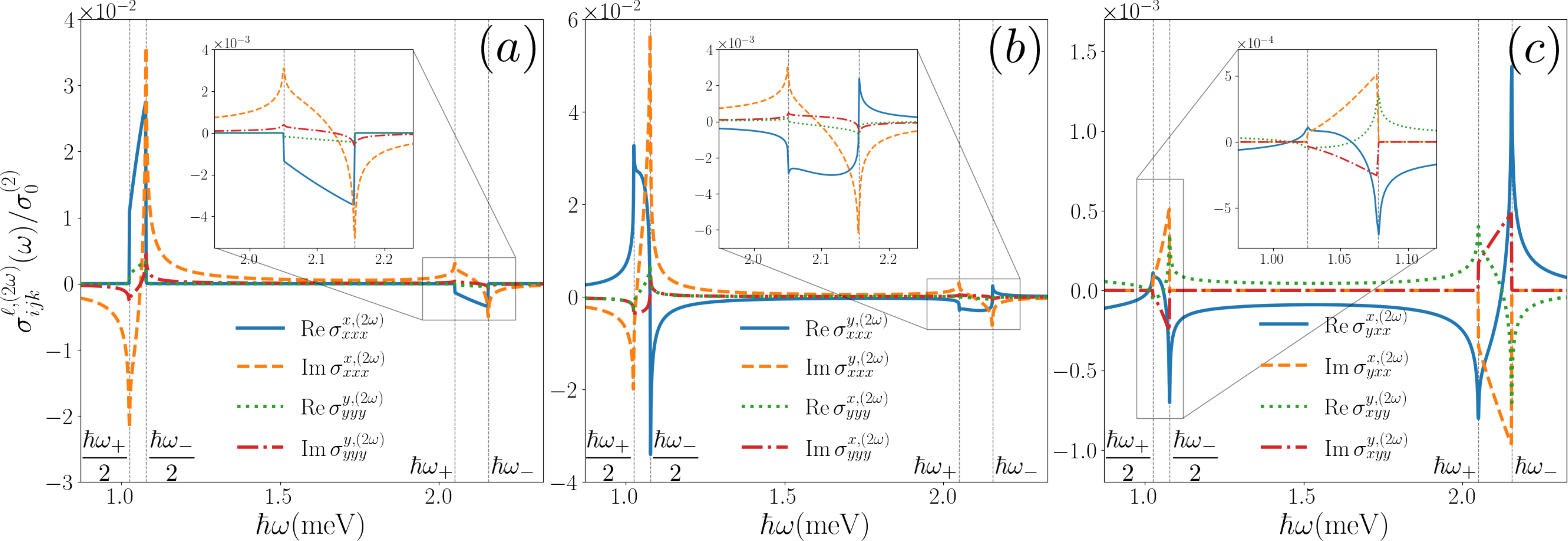}
\caption{Second-harmonic spin current conductivity tensor $\sigma^{\ell,(2\omega)}_{ijl}(\omega)$ for a gapped 2D anisotropic Rashba model, which determines a spin current flowing in direction $i$, with the spin polarized along the direction $\ell$, induced by
the external field components $j$ and $l$. In this system $\sigma^{z,(2\omega)}_{ijl}(\omega)=0$.
(a) Longitudinal components with $i=j=l=\ell$. (b) Longitudinal components with $i=j=l\neq\ell$.
(c) Hall components with $i\neq j=l=\ell$.
The parameters used are the same as in Fig.\,\ref{2OMEGATWOLEVEL},
and $\sigma^{(2)}_{0} = e^2/4\pi\varepsilon_R k_R$.}
\label{fig:SPINEF2DELTA}
\end{figure}

A number of conclusions can be derived from these expressions.
In contrast to the linear response, the above expressions imply that there is no 
$z$-polarized SH spin current induced in the system, $\sigma^{z,(2\omega)}_{ijl}(\omega)=0$. On the other hand,
equation (\ref{SHiii}) shows that if $\Delta=0$ the longitudinal response $\sigma^{\ell,(2\omega)}_{iii}$ describes a spin current with the spin oriented 
perpendicularly to the electric field (and to the spin current), 
while one with parallel spin orientation and current is possible when $\Delta\neq 0$.
The Hall components $\sigma^{\ell,(2\omega)}_{ijj}$ (\ref{SHijj}) generate spin currents with spin always oriented in the direction of the field (normal to the spin current), regardless of the presence of an energy gap. As for the components in (\ref{SHiji}), 
$\sigma^{\ell,(2\omega)}_{iij}(\omega)$ and $\sigma^{\ell,(2\omega)}_{iji}(\omega)$, 
they are associated to a nonlinear current with spin orientation parallel 
to it when $\Delta = 0$, otherwise such an orientation is not fixed
for $\Delta\neq 0$.

The terms in braces in (\ref{SHijj})-(\ref{SHiji}) involve the masses through the  combination $m_d=\sqrt{m_xm_y}$ only. Thus,
the in-plane anisotropy becomes apparent in every non-zero tensor component through a factor of the type $ (m_x / m_y)^{\pm \nu/4}$, with $\,\nu=1$ or $3$. 
Figure \ref{fig:SPINEF2DELTA} shows the frequency dependence of the in-plane polarized SH components
$\sigma^{x,(2\omega)}_{ijl}(\omega)$ and $\sigma^{y,(2\omega)}_{ijl}(\omega)$.
As expected, the spectral structure around $\hbar\omega_{\pm}$ is now accompanied
by new features around the subharmonics $\hbar\omega_{\pm}/2$. The overall
structure can be modified through Fermi energy variation, given the behavior of the
functions $\hbar\omega_{\pm}(\varepsilon_F)$ observed in Fig.\,\ref{fig:JDOSCOLORTICKS}.

The SH spin current response (\ref{SHijj})-(\ref{SHiji}) presents characteristic differences with respect to the linear response, as is the case in the model of section III. Diagonal components are present even in absence of an energy gap, subharmonic structure appears in the spectrum, and the spin polarization is no longer 
out-of-plane, so that a nonlinear spin Hall effect with in-plane spin orientation 
is generated. These features may be useful in nonlinear optical spintronic devices.

\section{Summary}

 We calculated the spin conductivity tensors which characterize the 
 electric-dipole induced spin currents at the fundamental and second harmonic
 frequencies, in two anisotropic systems with SO interaction. 

 In the case of a 2DEG with R+D$[hkl]$, a time-reversal preserving system,
 the anisotropy arises from the 
 interplay between the Rashba and Dresselhaus couplings, which in turn depends
 sensitively on the sample growth direction. For a given crystallographic 
 orientation, the spin splitting of the states acquire a particular dependence on
 the direction in ${\bf k}$-space. This modify the spectrum of allowed optical
 transitions, and the JDOS and the spin current responses at $\omega$ and $2\omega$
 display characteristic spectra. This suggests an additional way to influence
 the linear and nonlinear spectra by choosing in advance the growth direction
 of the sample, besides frequency tuning, the modulability of the
 Rashba strength, or the direction of the applied electric field. 
 We found also that the response functions
 $\sigma^{\ell}_{ij}(\omega)$ and $\sigma^{\ell,(2\omega)}_{ijl}(\omega)$
 vanish identically under the SU(2) symmetry conditions found in Ref\,\cite{Kammermeier2016a}. There are, however, additional conditions
 under which specific tensors components vanish, without the requirement
 of having a collinear SO vector field. Thus, by a proper choice of the
 growth direction and SO material parameters, one could select the
 polarization of the linear and SH spin currents according to the direction of 
 flowing.

 In the case of the anisotropic Rashba model studied in Sec.\,\ref{ARM}, the
 anisotropy is that of a 2D free electron gas with different masses \cite{Ahn2021a}, $m_x\neq m_y$,
 in the presence of a Rashba type interaction which introduces different spin splitting
 along perpendicular directions \cite{Popovic2015}. To be comprehensive, the model includes an energy gap parameter, which breaks the time-reversal symmetry; when the gap is closed, the model reduces to that studied in
 Ref.\,\cite{Saberi-Pouya2017}.
 The band structure offers distinct positions for the Fermi level
 (above, within, and below the gap), which define
 several distinguishable scenarios for the allowed optical interband transitions,
 characterized by the Fermi contours in each case. These manifest in contrasting
 ways in the linear and SH spin current response. The linear spin conductivity
 $\sigma^{\ell}_{ij}(\omega)$ shows that only out-of-plane spin polarized currents develops ($\ell=z$), while the SH spin conductivity tensor $\sigma^{\ell,(2\omega)}_{ijl}(\omega)$
 gives rise to currents with spin orientation lying parallel to the plane of the electron gas ($\ell=x,y$). The longitudinal components $\sigma^{\ell}_{ii}(\omega)$
 are inversely proportional to $\sqrt{m_i}$, connected through the masses in the form
 $m_x\sigma^z_{xx}(\omega)=m_y\sigma^z_{yy}(\omega)$, and vanishing for the gapless case.
 In contrast, the Hall components are non null regardless of the presence of a
 gap, depend on the masses through the geometric mean $\sqrt{m_xm_y}$ only, and
 satisfy $\sigma^{\ell}_{yx}(\omega) = - \sigma^{\ell}_{xy}(\omega)$. On the other
 hand, the SH components are proportional to the ratio of masses in the combination
 $(m_x/m_y)^{\pm 1/4}$ or $(m_x/m_y)^{\pm 3/4}$.

 In summary, we investigated the spectral properties of the
 linear and nonlinear optical spin conductivities of two anisotropic 
 models for SO coupled systems, and its dependence on a number of
 physical quantities like the exciting frequency, the position of the Fermi level,
 energy gap, mass anisotropy, SO strengths, Rashba and Dresselhaus
 couplings interplay for arbitrary sample growth directions, or
 the direction of the externally applied electric field, according to 
 each case. The presence of anisotropy introduces optical signatures 
 which in turn may be useful to identify or estimate some of these 
 material parameters.  
 In particular, the models illustrate the existence of the 
 nonlinear spin Hall effect in systems with SO interaction, under the presence or 
 absence of time-reversal symmetry. 
 The results suggest different ways 
 to manipulate the optically induced linear and SH spin current responses,
 which could find spintronic applications.
We hope that this work will stimulate further investigations under more general conditions, such as the presence of extrinsic SO mechanisms or
the use of a conserved spin current definition \cite{Chen2014,Shi2006}.

	\vspace{1cm}
	\bibliographystyle{apsrev4-2}
	\bibliography{main1}
	
\end{document}